\documentstyle[aaspp4,11pt]{article}

\slugcomment{Accepted for publication in the ApJ}

\lefthead{Matsumoto et al.}
\righthead{X-Ray Properties of Early Type Galaxies}

\singlespace

\begin{document}

\title{X-Ray Properties of Early Type Galaxies as Observed with 
{\it ASCA}}

\author{H. Matsumoto\altaffilmark{1}, K. Koyama, H. Awaki, and T. Tsuru}
\affil{Department of Physics, Kyoto University, Sakyo-ku, Kyoto, Japan 606-01}

\author{M. Loewenstein}
\affil{Laboratory for High Energy Astrophysics, NASA/GSFC, Code 662,
Greenbelt, MD 20771}

\and
\author{K. Matsushita}
\affil{Department of Physics, University of Tokyo, 7-3-1 Hongo, Bunkyo-ku, 
Tokyo, Japan 113}

\altaffiltext{1}{E-mail address: matumoto@cr.scphys.kyoto-u.ac.jp}

\begin{abstract} 
We have systematically investigated the {\it ASCA} spectra of 12 early 
type galaxies. 
This paper presents the global spectral properties of these 
systems based on a larger sample than in any previous {\it ASCA} study. 
The X-ray spectra were uniformly fitted by a two-component model consisting 
of hard X-rays from thermal emission with a temperature of about 10 keV
or from a power-law with index 1.8, 
plus soft X-rays from a thin thermal plasma with  
temperature ranging from 0.3 to 1 keV. 
The X-ray luminosities of the hard component are found to be
proportional to the blue band luminosities, 
while those of the soft component show large scatter with 
no clear correlation. 
 The metal abundances determined from the soft component are 
systematically lower than solar, with a mean value of about 0.3 solar. 
We examine the relationships between 
the temperature and volume emission measure, 
and between the gas temperature and the stellar velocity dispersion. 
The volume emission measures for  
early type galaxies  plotted as a function of the gas temperature are 
well below the extrapolated line found in clusters of galaxies, indicating 
that early type galaxies
are relatively gas poor compared with galaxy clusters. 
 The ratio of the stellar kinetic energy per unit mass 
to the thermal energy of the hot gas per unit mass ($\beta_{spec}$) 
is less than unity. 
We found no systematic relationship between X-ray properties
and environment, suggesting that the interaction between  
interstellar matter and the intracluster medium is not strong. 
\end{abstract}


\section{Introduction}

The {\it Einstein} satellite discovered X-ray emitting gas halos  
around luminous early type  galaxies (eg. Forman, Jones, \& Tucker 1985 
\markcite{FJT85}),  
forcing the rejection of previous assumptions that elliptical galaxies  
are gas poor systems.  
The X-ray surface brightness distribution was found to
closely follow the optical image 
(Trinchieri, Fabbiano, \& Canizares \markcite{TFC86} 1986); 
however the X-ray  luminosity, and hence the mass of the hot gas, 
showed large scatter from galaxy to galaxy  (eg. Canizares, Fabbiano,
\& Trinchieri \markcite{CFT87} 1987).   
The estimated cooling time of the hot gas is less 
than the Hubble time (Trinchieri et al. \markcite{TFC86} 1986),  
however cooling may be balanced 
by heating mechanisms such as supernova explosions. 
 Arnaud et al. \markcite{Arn92} (1992) found that the iron mass 
in the intracluster medium (ICM) is directly proportional to 
the optical luminosity from early type galaxies in clusters of galaxies. 
This leads one to suspect that metals in the ICM originate in  
early type galaxies and that there might be a link, in the form
of early star formation and galactic winds, between
the ICM and metal abundances in early type galaxies.
Due to the poor energy resolution of previous satellites,  however,  
critical parameters for investigating the nature of the hot gas
in early type galaxies, 
such as the gas  temperature, mass, and metal abundance  have not 
been  well-constrained.   
Accordingly,  several important issues including 
the origin,  heating, and fate of the hot gas, and 
the effect of early type galaxies on the ICM, 
are not well understood. 

{\it ASCA} (Tanaka et al. \markcite{TIH94} 1994) is the first 
satellite with sufficient energy 
resolution and large effective area over the broad 
0.4 to 10 keV bandpass to determine these physical parameters. 
With this satellite, Awaki et al. \markcite{Awaki94} (1994) 
analyzed 3 early type galaxies 
and found that the temperature and metal abundance of 
the hot gas are about 0.8 keV and 0.4 solar, respectively. 
The inferred metal abundance is lower than expected
from theoretical  models, and lower than those  determined at  
other wavelengths.  
Accordingly,  the {\it ASCA} observations  raise a  serious 
challenge to standard models for the  evolution  of galaxies 
and clusters of galaxies. 
This mystery deepened with the discovery of even lower 
metallicities (less than 0.2 solar) in two other elliptical 
galaxies (Loewenstein et al. \markcite{LMT94} 1994), 
and in the outer regions of NGC4636 (Mushotzky et al. 
\markcite{MLA94} 1994).
Matsushita et al. \markcite{Kyoko94} (1994) discovered,  
analyzing 5 early type galaxies including the sample of 3  
in Awaki et al. \markcite{Awaki94}(1994), 
a hard X-ray component at energies above a few keV  
in addition to the hot gas component dominating below about 2 keV.   
In order to  perform plasma diagnostics of the hot gas component,  
it is essentially  important to estimate  and remove 
the hard component from the X-ray spectra of early type galaxies. 
If the hard  component is due to the integrated emission 
of stellar sources,  as suggested by Matsushita et al. 
\markcite{Kyoko94} (1994),  its luminosity should be proportional 
to optical luminosity. 
    Since the samples in previous {\it ASCA} studies are too limited to 
fully investigate all the issues mentioned above,   
we have systematically  studied twelve early type galaxies, 
including the samples of Awaki et al. \markcite{Awaki94}(1994) and 
Matsushita et al. \markcite{Kyoko94} (1994) 
as well as newly observed galaxies. 
These comprise the largest sample of {\it ASCA} spectra of early 
type galaxies yet investigated.

\section{Observations and Data Reduction}

The galaxies we have selected for this study are   
NGC4472, NGC4406, NGC4636, NGC4649, NGC499, NGC507, NGC720, 
NGC4374, IC4296, NGC4382, IC1459 and NGC4365, 
which are listed  in Table 1 with their optical parameters.  
The  sample, which includes 7 performance and verification (PV) phase
targets and 3 first-round guest investigator observations 
by the first author (HM), was principally selected on the basis 
of soft X-ray fluxes measured with  the {\it Einstein} satellite 
in order to obtain useful spectra in reasonable exposure times. 
In addition to this criterion,  a number  of 
low X-ray luminosity galaxies were intentionally
selected in order to  detect and study possible  
differences between gas-rich  and  gas-poor galaxies.    
Furthermore, we selected our sample to include both galaxies 
in dense and tenuous gaseous environments in order to be able to 
study possible interaction between the ICM and interstellar medium (ISM).

  All the galaxies were observed with two solid-state imaging 
spectrometers (SIS0 and SIS1) and two gas imaging spectrometers 
(GIS2 and GIS3) at the foci of 4 thin-foil X-ray mirrors (XRT) on board 
the {\it ASCA} satellite. Details of the instruments can be found in 
Burke et al. \markcite{BMH91} (1991), Ohashi et al. \markcite{Ohashi96a} 
(1996a), Makishima et al. \markcite{Max96} (1996), and 
Serlemitsos et al. \markcite{Ser95} (1995), 
while a general description of {\it ASCA} can be 
found in Tanaka et al. \markcite{TIH94} (1994). 
In table 2, we summarize the {\it ASCA} observation log of 
our sample galaxies. 
The data were screened with the standard selection criteria: 
data affected by the South Atlantic Anomaly, 
Earth occultation, and regions of low geomagnetic rigidity were excluded. 
We also eliminated contamination by the bright Earth, 
removed hot and flickering pixels for the SIS data, 
and applied rise-time rejection to exclude particle events for the GIS data. 
After the above data screening, we extracted X-ray spectra from circular 
regions centered on each galaxy with a uniform physical
extraction radius of 38 kpc -- the largest radius completely
contained within the SIS field-of-view for all galaxies.
The background spectra  were obtained from source free regions 
around each galaxy.

\section{Spectral Analysis and Results}

In order to utilize the  $\chi^2$ technique, 
we rebinned the X-ray spectra to contain at least 20 counts 
in each spectral bin, then simultaneously fitted the spectra 
from the SIS0, SIS1, GIS2 and GIS3,   
with a common normalization factor 
but separate response functions for each detector,  
using the XSPEC (version 8.50) spectral fitting package. 

Awaki et al. \markcite{Awaki94} (1994) reported that  
the SIS spectra of bright early type galaxies, 
NGC4636, NGC4472 and NGC4406 exhibit emission lines 
from highly ionized Fe, Si, and S atoms.  
We further confirmed the presence of these emission lines 
in the spectra of the additional bright galaxies  
NGC4649, NGC499, NGC507 and NGC4374. 
Therefore we inferred  that thin hot plasma emission is prevalent 
in bright early type galaxies and, hence, attempted to fit the X-ray 
spectra with a single temperature thermal plasma model 
(Raymond \& Smith \markcite{RS77} 1977; hereafter RS model) 
modified by interstellar absorption. 
The free parameters were the plasma temperature, 
metal abundance, photoelectric absorption column density, 
and normalization.  
The abundances are collectively varied keeping the relative ratio at 
the solar value (Anders \& Grevesse \markcite{AG89} 1989). 

This model, however, failed to reproduce the observed spectra except 
that of NGC499.   From all galaxies,  systematic excess  
emission is found above the 4 keV band.  
A typical example, NGC4649, is shown in figure 1. 
Thus we  found  the presence  of hard X-ray emission, 
as previously reported in several individual galaxies 
(Kim, Fabbiano \& Trinchieri \markcite{KFT92} 1992; Matsushita et al. 
\markcite{Kyoko94} 1994),  
to be a ubiquitous property of elliptical galaxies.  
In figure 2 we compare the X-ray spectra of the two elliptical galaxies 
NGC4636 and NGC4365; 
although they have nearly identical optical blue luminosities, 
the former is X-ray bright and  the latter is  X-ray faint. 
   The difference in X-ray flux is confined to the soft X-ray band:  
the X-ray spectrum of NGC4636  below 3 keV is  
nearly two orders of magnitude brighter than that of NGC4365. 
On the other hand,  the X-ray spectra above the 4keV band are 
indistinguishable, as might be expected based on 
the identical optical blue band magnitudes. 
Thus we compared the X-ray spectra above the 4 keV band for 
the entire sample,  and found no evidence of significant differences in 
spectral shape, although the absolute fluxes vary  
from galaxy to galaxy.   
Accordingly, we  assumed a  thermal bremsstrahlung 
model for the hard component to estimate the temperature variation  
using the data above 4 keV.  
Since the statistics are limited,  however, 
we could only determine a lower limit on the temperature 
of the bremsstrahlung model of  about 2 keV. 
Since the GIS has higher sensitivity than the SIS 
in the relevant high energy band, 
we co-added all of the GIS spectra in our sample to improve 
the statistical accuracy.  
Figure 3 shows the composite GIS 2 and 3 spectra above the 4 keV band. 
    We fitted  a  thermal bremsstrahlung model to 
this  spectrum, and obtained an acceptable fit ($\chi^2/d.o.f.$ of 
$22.36/31$) with the best-fit temperature $12.0 ^{+29.3}_{-5.5}$ keV 
(hereafter, errors are 90 \% confidence unless otherwise stated). 
  This composite spectrum is equally well  fitted 
($\chi^2/d.o.f.$ of 22.83/31) 
by a power law model 
with photon index 1.8$\pm0.4$. 
 Therefore, we cannot distinguish between the two possible origins -- 
thermal and non-thermal -- of the hard component  
from the spectrum alone.
 However the  relevant physical parameters, such as the 
X-ray luminosity ($L_X$) of the hard component, or 
iron K-shell line flux  do not 
significantly differ whether
we use thermal bremsstrahlung or power-law models. 
Thus we assume the thermal-bremsstrahlung model for further discussion.

In order to estimate the strength of Fe K line emission,  
we added a Gaussian line to the above model.    
The best-fit temperature is slightly altered to $8.8^{+13.9}_{-3.7}$ keV, 
with line center energy and equivalent width 
of the Gaussian of $6.4^{+0.3}_{-0.4}$ keV and $220(<460)$ eV, 
respectively. 
Confidence contours of the line energy 
and equivalent width at $\Delta\chi^2~=$ 2.71, 4.61 
and 9.21 are plotted in Figure 4. 
 Adding the line decreased $\chi^2$ from 22.36 to 18.50, 
which is not statistically significant (Malina, Lampton \& Bowyer 
\markcite{MLB76} 1976). 
Therefore we can conclude that the composite hard component
spectrum requires no significant emission lines.  

  Since the presence of both emission lines from
highly ionized heavy elements in the low energy band 
and a hard X-ray tail at the high energy  band seem 
to be common in the galaxies in our sample, 
we fitted two-component models to all broad band spectra. 
The models consist of a thermal bremsstrahlung component   
with a fixed temperature of 12 keV 
determined from the composite hard band spectrum, 
plus a thin thermal model (RS model) for the soft band emission. 
The column density of the hard component was constrained to 
be the same as that of the RS model.
The free parameters were the temperature and metal abundance of the
RS model, the column density, and normalizations for both  
the RS and bremsstrahlung models. 
The best-fit parameters and  90\% confidence errors 
are listed in Table 3. 
We generally found reasonable fits with this unified model, 
however two galaxies --  NGC4636 and NGC507 -- exhibit unacceptably 
large values of $\chi^2$. 
Although these two galaxies may require more complicated modeling
(NGC4636 was found to have both an abundance gradient and an 
extra centrally concentrated cool component by Mushotzky et al. 
\markcite{MLA94} 1994),
more detailed study of individual galaxies is beyond 
the scope of this paper. 
The best-fit RS abundances and temperatures are in excellent 
agreement with those derived from {\it ROSAT} PSPC observations 
(Davis \& White \markcite{DW96} 1996; 
Kim \& Fabbiano \markcite{KF95} 1995), 
except for NGC4472 where the {\it ASCA} parameters 
are consistent only with the central PSPC parameters 
(Forman et al. 1993 \markcite{FJD93}). Significant
excess absorption is required by the {\it ASCA}, 
but not the PSPC, spectra of NGC4472, NGC4406, and NGC4649. 
This discrepancy is likely due either to remaining inaccuracies 
in the {\it ASCA} low energy response, 
or to the inadequacy of the simple foreground screen model 
that we employ for the absorption.

Many authors have pointed out systematic differences 
in the best-fit parameters between models using different 
plasma emission codes.  
We thus examined this possible systematic and model-dependent uncertainty 
by comparing the RS and MEKA models (Mewe, R., Gronenschild, E.H.B.M., 
and van den Oord, G.H.J. \markcite{MGO85} 1985). 
In figures 5a -- d, we show the correlation plots of best-fit parameters
between the RS and MEKA models: the column density (a), temperature (b),
metal abundance (c), and X-ray luminosity (d).
 The best-fit column densities are found to be consistent within the errors,  
while best-fit temperatures in the  MEKA model are almost 20 \% 
lower than the RS model.
 The metal abundances in the MEKA model generally have larger values 
than the RS model, although the scatter in the correlation plot is large. 
As for the luminosity, both models give essentially the same best-fit value. 
 Thus we confirmed the existence of plasma model-dependent systematic 
and non-systematic errors. 
However these do not largely effect the following discussion, 
hence we refer to the best-fit values using the RS model.

\section{Discussion}

\subsection{The Hard Component}

Figure 6 shows the relation between the X-ray luminosity ($L_X$) 
of the hard component in the 0.5--4.5 keV band and 
the blue band luminosity ($L_B$). 
The uncertainties in $L_X$ of the hard component are 
typically $\sim$30\%. 
The solid line is the best-fit relation  
between the blue luminosity and the 0.5--4.5 keV band luminosity 
found for bulge-dominated  spiral galaxies 
(Canizares et al. \markcite{CFT87} 1987). 
We found that $L_X$ of the hard component is proportional to $L_B$, 
and that the constant of proportionality is nearly the same as 
found in spiral galaxies:  $L_X/L_B$ for the hard component 
in early type galaxies is generally consistent with 
that of bulge-dominated spiral galaxies like M31. 
The excess  hard flux in the radio galaxies IC1459 and IC4296 
probably can be explained by X-ray emission from AGN; 
indeed our analysis of archival PSPC data from 
IC1459 indicates the presence of an unresolved hard component. 
Furthermore, we found that the composite {\it ASCA} GIS 
spectrum above the 4keV band (Section 3) resembles the X-ray spectra of 
bulge-dominated spiral galaxies which can be fitted with a thermal 
bremsstrahlung model with $kT~^>_\sim~5$ keV (eg. Makishima et al. 
\markcite{Max89} 1989). 

These facts suggest that early type galaxies 
contain a hard X-ray component attributable to 
the same origin as X-rays in bulge-dominated spiral galaxies.    
Since the X-ray emission from bulge-dominated spirals is thought 
to arise from a population of low-mass X-ray binaries (LMXB), 
we suspect that the primary hard component in early type galaxies 
is also due to the integrated emission of an ensemble of LMXB.  
Matsushita et al. \markcite{Kyoko94} (1994) came to the same conclusion
based on a more limited (5 galaxies) sample; they also discovered that
the hard component is generally extended which argues against an AGN power-law
origin for the hard component.
 With a larger galaxy sample than Matsushita et al. \markcite{Kyoko94} 
(1994), we therefore confirm that both
the shape and relative magnitude of the hard component in
early type galaxies is consistent with that in
bulge-dominated spiral galaxies, strengthening
the suggestion that the hard X-ray component in early 
type galaxies is due to LMXB. 

  Although LMXB usually exhibit a weak 6.7 keV iron line,
we found no significant iron line in the early type galaxies. 
The 90 \% confidence upper limit on 
the equivalent width of the Fe K$\alpha$ line of 500 eV is too large
to place 
any strong constraints on the LMXB scenario.

\subsection{The Soft Component}

Emission lines from highly ionized Fe, Si, and S are detected from 
the 7 galaxies in our sample with the best statistics; 
non-detection of emission lines for the other galaxies is consistent 
with their having a lower signal-to-noise ratio. 
The existence of these emission lines is direct evidence 
for the prevalence of thin hot interstellar plasma 
in bright early type galaxies. 
We now discuss the nature of this plasma using our results, 
as well as  referring to published  optical data.

\subsubsection{Correlation between $kT$ and $VEM$} 

Because $L_X$ of the hard component is generally 
proportional to $L_B$, 
the large dispersion in the total $L_X$-$L_B$ diagram 
(Canizares et al. \markcite{CFT87} 1987) is mainly due to 
a spread in $L_X$ of the soft component. 
This is illustrated 
in figure 7 where 
$L_X$ of the soft component is plotted versus $L_B$.
 Since, ignoring the mostly small variations in temperature,  
$L_X$ of the soft component is  proportional to 
the volume emission measure ($VEM$), and 
\begin{displaymath}
VEM \equiv {\int}n_e n_p dV \sim M_{gas}^2/V
\end{displaymath}
, where $n_e$ and $n_p$ are the densities of electrons and protons, 
respectively, and
$V$ and  $M_{gas}$ are, respectively, the volume and  mass of the hot gas, 
the large dispersion in $L_X$ found in figure 7 indicates the presence of a
wide mass range of hot gas in early type galaxies.

Figure 8 is a plot of the volume emission measure ($VEM$) versus  
the temperature of the hot gas.  
For comparison, we show the data for clusters of galaxies 
in the same figure with plus symbols (Hatsukade \markcite{Hatsu89} 1989). 
The Spearman rank correlation coefficient between $VEM$ and temperature for 
early type galaxies is estimated to be 0.43; therefore, we cannot conclude
that there is a strong correlation between 
temperature and mass of the hot gas. 
The line in the figure shows the best-fit relation for the cluster data. 
 What is clear is the fact that most of our galaxy sample is 
located well  below the cluster correlation line, and that $VEM$ drops 
steeply with $kT$ from rich cluster all the way down to galaxy scales.
Since for 
a constant gas mass as a fraction of gravitational mass
$VEM$ $\propto kT^{1.5}$, it is clear that
early type galaxies are far more gas poor than clusters.

\subsubsection{Heating Mechanism}

Figure 9 presents the relation between the stellar velocity dispersion and 
the hot gas temperature.  
The velocity dispersions of the galaxies in clusters are shown 
with plus signs (Hatsukade \markcite{Hatsu89} 1989). 
The solid line represents $\beta_{spec} = 1$, where 
\begin{displaymath} 
\beta_{spec} \equiv 
\frac{\displaystyle \mu m_H \sigma^2}{\displaystyle  kT}
\end{displaymath}
, and $\mu$ , $m_H$, $\sigma$, $k$ and $T$ are 
mean molecular weight (we assumed it to be 0.6), proton mass, stellar 
(or galaxy) velocity dispersion, Boltzmann constant, and temperature, 
respectively. 
$\beta_{spec} = 1$ 
indicates that the thermal energy of the hot gas per unit mass 
is equal to the stellar (or galactic) kinetic energy per unit mass. 
The Spearman rank correlation coefficient for early type galaxies between
 $\sigma$ and temperature is estimated to be 0.54,
 which is not significant.
 Thus we cannot strongly conclude that 
there is a positive correlation between 
these two parameters which would indicate a correlation
between the depth of the gravitational potential and 
the temperature of the hot gas as expected for equilibrium systems.
 However, all our sample galaxies show
higher temperatures than would be expected based on
the best-fit cluster sample correlation line; {\it i.e.}
$\beta_{spec}$ is less than 1 
for all of the early type galaxies in our sample compared to
approximately 1 for clusters of galaxies. 
In other words, the thermal energy of the hot gas per unit mass 
is larger than the stellar kinetic energy per unit mass. 

Figure 10 shows the relation between the velocity dispersion and 
$\beta_{spec}$. The mean value of $\beta_{spec}$, weighted using 
relative errors, is about 0.5. 
If the stars and gas in early type galaxies are both 
isothermal and in hydrostatic
equilibrium, $\rho_{gas}\propto \rho_{gal}^{\beta_{spec}}$, 
where $\rho_{gas}$ and $\rho_{gal}$ are the densities 
of hot gas and stars in the galaxy, respectively (eg. Sarazin 
\markcite{Sarazin86} 1986). 
Trinchieri et al. \markcite{TFC86} (1986) have found that 
for bright early type galaxies, 
the X-ray and optical surface brightness tend to follow each other. 
Since the X-ray brightness profile traces 
$\rho_{gas}^2$ and the optical profile traces $\rho_{gal}$, 
this implies that $\beta_{spec}$ is about 0.5 (Fabbiano 1989 
\markcite{Fabbiano89}),
consistent with our direct measurement of $\beta_{spec}$. 
$\beta_{spec}=0.5$ is expected if cooling (proportional to 
$\rho_{gas}^2$) is approximately 
balanced by heating by stellar winds, gravitational compression in 
an inflow, or Type Ia supernovae that all scale 
as the local stellar density.

Assuming that the present-day 
stellar mass loss rate is about $1.5 \times 10^{-11}~L_B 
M_{\sun}/yr$ and has decreased with time as $t^{-1.3}$ 
(Ciotti et al. \markcite{CDP91} 1991), we see that the typical 
mass of hot gas observed in early type galaxies 
($10^{9-10}~M_{\sun}$) is much less than the total accumulated 
over the Hubble time ($\sim 10^{11} M_{\sun}$).  
This fact in conjunction with $\beta_{spec}=0.5$ and 
the short cooling times of the hot gas in the central 
regions ($\sim 10^6$ -- 10$^8$ yr; Fabbiano \markcite{Fabbiano89} 1989) 
implies that either (1) there is some heating mechanism that 
balances cooling and has driven much of the stellar mass loss out 
of the galaxy, or (2) an inflow is occurring where the primary 
heating mechanism is gravitational compression and 
where there must be some sink for gas 
cooling below X-ray emitting temperatures. 

 One may argue that a plausible heating mechanism is  
type Ia supernova explosions. 
We consider a simple model where the energy of supernova explosions 
is constant ($\sim 10^{51}$ erg) and the energy 
deposited per unit gas mass uniform over the entire galaxy. 
Let $k{\Delta}T$ be the ``excess temperature''
-- the difference between the temperature of the hot gas 
and that corresponding to the stellar velocity dispersion, 
\begin{displaymath}
k{\Delta}T \equiv kT_{{\rm hot~ gas}} - \mu m_H \sigma^2.
\end{displaymath} 
Since we are considering emission from a
standard volume, the mass of the hot gas is proportional 
to the square root of 
$VEM$ (Section 4.1.1), and the supernova rate 
thought to be proportional to $L_B$. 
If heating by type Ia supernovae is important, 
we would expect a negative correlation between $k{\Delta}T$ and 
$VEM^{1/2}/L_B$ because the rise in gas temperature should be smaller 
as the gas mass per supernova becomes larger. 

Figure 11 shows the plot of $VEM^{1/2}/L_B$ vs. $k{\Delta}T$. 
Since the Spearman rank correlation coefficient is 0.40, 
we can conclude that there is no negative correlation. 
This contradicts simple models where 
the heating source is proportional to $L_B$ 
and the efficiency of heating is uniformly high, 
but rather suggests that the efficiency of energy transfer 
to the hot gas may increase as the gas mass per star becomes larger. 
A negative correlation 
between $VEM^{1/2}/L_B$ and Fe abundance would also be predicted by
a simple SNIa-dominated heating model
(see below), and is likewise not seen. 

The importance of heating by type Ia 
supernovae can generally be ruled out on the basis 
of the observed constraints on the iron abundance, since each  
explosion injects $\sim 0.7~M_{\sun}$ of Fe into 
the hot gas. This means that increasing the gas 
temperature by $\sim 0.5$ keV will enhance the Fe abundance 
by $\sim 1$ in solar units. Even if we attribute all of the 
Fe in the hot gas to supernova explosions -- an extreme 
assumption considering the expected contribution from 
stellar mass loss (Section 4.2.3) -- supernova heating cannot 
explain the observed values of $k{\Delta}T$ in most of our sample.

\subsubsection{Metal Abundance}

Figure 12 shows the relation between the abundance and the temperature. 
We can see no clear correlation between these parameters, although 
the lowest metallicities are found in the galaxies with 
the coolest ISM. 
If we use the MEKA model, these conclusions are not essentially 
altered. 

For the galaxies whose metal abundances are well determined, 
the abundances, without exception, are found to be less than half solar. 
These values are considerably lower than theoretical expectations,  
and those determined at optical wavelengths. 
An abundance gradient (Mushotzky et al. \markcite{MLA94} 1994) 
may partially account for the discrepancy between optical abundances, 
which are determined in the central region of the galaxy, 
and X-ray abundances measured over a region extending 
to many effective radii. 

Significantly sub-solar metallicities have also been observed 
with {\it ASCA} in a variety of sources, including 
stellar coronae (eg. Singh, Drake, \& White \markcite{SDW95}
1995), starburst galaxies (eg. Awaki et al. \markcite{Awaki95} 1995; 
Tsuru et al. \markcite{TGT96} 1996), 
and supernova remnants (SNRs) (eg. Hayashi et al. \markcite{Hayashi96} 1996). 
We think it is worth noting that 
some SNRs in the Large Magellanic Cloud sufficiently evolved  
to sweep up large amounts of ISM exhibit X-ray spectra which closely 
resemble those of early type galaxies 
(Hayashi et al. \markcite{Hayashi96} 1996). 
Improvements in our understanding of the processes of enrichment by 
supernovae and stellar mass loss  
may help solve the puzzle of low abundances if, for example, 
the gas ejected from supernovae does not mix well into the ISM 
and the gas which is overly dense and metal-rich cools and 
drops out of the X-ray emitting hot gas 
(Fujita, Fukumoto, \& Okoshi \markcite{FFO96} 1996). 
 
Arimoto et al. \markcite{Ari96} (1996) suggest that low abundances 
may be an artifact, 
since the metal abundance is constrained by Fe L 
emission lines around 1 keV for which the atomic physics parameters 
are poorly known. 
However, Hwang et al. \markcite{HML96} (1996) have found that, 
for hot gas temperatures between 2 and 4 keV,
the abundance determined from Fe L and Fe K lines
is fairly consistent (with abundances determined
from the Fe L region actually slightly higher), and
any errors in the abundances due to atomic 
physics uncertainties that exist in the current codes 
are unlikely to exceed  30 -- 50 \%.
Therefore plasma model uncertainties may not explain 
the low abundances, although the increasing complexity of 
Fe L emission for temperatures below 1 keV imply 
that more detailed analysis than lies within the scope of this paper 
is needed to solve this problem. We note that the consistency of 
Fe abundances derived independently over 
a wide range of ionization states in stellar coronae 
(eg. Drake \markcite{Drake96} 1996) 
and the comparably low abundances inferred for ions 
with more straightforward atomic parameters 
(eg. Si; Ohashi et al. \markcite{Ohashi96b} 1996b)
implies that it is unlikely that the abundances derived 
here are underestimated by large factors. 

Weak constraints on oxygen abundances are  
available only for the 3-brightest  early type galaxies 
(NGC4472, NGC4406, and NGC4636); other galaxies have
insufficient signal-to-noise ratios. 
The best fit oxygen abundances for the 3 galaxies we derive are  
consistent with those found by Awaki et al. 
\markcite{Awaki94} (1994), roughly 
equal to the iron abundances.
The oxygen abundances are not determined to sufficient accuracy to
place significant constraints on the relative contributions of SNII
and SNIa to the heavy metal production in the observed galaxies.

\subsubsection{Interaction between ISM and ICM}

Our sample includes galaxies in both relatively dense and tenuous 
intergalactic environments.
NGC4636, for example, is very far from the center 
of the Virgo cluster and no nearby ICM X-ray emission is seen, 
while NGC4472, the central galaxy of a
subgroup in the Virgo cluster, is surrounded by significant ICM X-ray
emission. 
NGC720, at the center of the poor group LGG38 
(Garcia \markcite{Gal93} 1993), on the other hand,
does not show effects from either ram pressure or tidal distortion, hence 
is regarded as rather isolated 
(Buote \& Canizares \markcite{BC96} 1996). 

For all the correlations shown in figures 6 -- 12, 
we see no systematic differences between the galaxies in dense 
and tenuous gaseous environments.
This indicates that the interaction between the 
ICM and ISM is not strong enough to manifest itself in our sample. 
More comprehensive  investigations  of 
such interactions require detailed study of spatial 
variations in  X-ray spectral properties as in Iwasawa et al. 
\markcite{IWF96} (1996) who have 
found evidence for ram pressure stripping of the ISM in NGC4406, as well
as a larger sample spanning a wider range of environments.

\section{Conclusion}

We have presented results from the uniform analysis of the {\it ASCA} 
spectra of 12 early type galaxies. 
This is the largest such sample to date and, as summarized below,
we have uncovered new facts as well as confirmed previous 
inferences based on much smaller samples.

\begin{enumerate}
\item

We found that the X-ray spectra from the 12 galaxies in our sample 
generally consist of at least two components: 
a soft component from X-ray emission of hot gas,
and a hard component that 
can be characterized by either a thermal 
bremsstrahlung model with $kT>5$ keV or a power-law model of 
photon index 1.8. 
The consistency of the strength and spectral shape of the hard component 
with those in bulge-dominated spirals corroborates previous 
suggestions that it 
primarily originates as the integrated emission from X-ray binaries. 
The hard component spectrum requires no significant Fe K line emission, 
and the upper limit on the equivalent width of Fe K line emission 
is 460 eV. 

\item
We confirmed that the metal abundance of the hot gas is generally 
less than 0.5 solar. 
There is no clear correlation between the abundance and temperature 
of the hot gas, however the lowest abundances are found in the
coolest galaxies.

\item
There is a large dispersion in the ratio of gas mass ({\it i.e.}
the X-ray luminosity of the soft component) to optical luminosity, 
implying a wide variety of gas-dynamical histories. 

\item
Theories of large-scale structure imply
that there is a continuum of dark matter halos spanning galaxy
and cluster scales. However, we have found that 
early type galaxies fall well below 
the constant gas mass fraction extension 
of the hottest clusters in the $kT-VEM$ plane (VEM $\propto kT^{1.5}$),
and that the ratio of thermal to kinetic energy 
per unit mass is, on average, twice that found for clusters of
galaxies.
This suggests that larger systems are generally more efficient 
at retaining their hot gas and less efficient at converting 
gas into stars. The interstellar media of elliptical galaxies 
are not simply scaled down version of intracluster media, but 
have been more profoundly effected by feedback from star formation 
and other gas-dynamical processes. 

\item
The low Fe abundances, as well as
the absence of a negative correlation between
the ``excess temperature'', defined as the difference between 
the hot gas temperature and the temperature corresponding to the stellar 
velocity dispersion, 
and the ratio of the square root of the volume emission measure of 
the hot gas to the blue band luminosity (an indicator of 
the mass ratio of hot gas to stars) argue against 
type Ia supernovae being an important heat source for the ISM.

\item
We found no systematic differences in the physical parameters of 
the hot gas between galaxies in dense 
and tenuous intracluster environments, although our ability
to test this is limited. 
This indicates that interaction between the intracluster medium and 
interstellar matter may not be strong. 
\end{enumerate}

In short, {\it ASCA} analysis of early type galaxies 
has yielded both expected
(a hard component with the same normalization as in spiral bulges) and
unexpected (low interstellar abundances) results.
The interrelationships and dispersion among the properties of the
hot gas component, 
and their comparison with those in intracluster 
media demonstrates the relative importance of astration, mass loss, and
gas-dynamics in determining the evolution of early type galaxies and their 
ISM. Any complete theory of galaxy formation must explain 
the observed trends and diversity of bound distributions of 
hot gas on galaxy 
and cluster scales.

\vspace*{1em}

\acknowledgements
We thank all of the members of the {\it ASCA} team and the launching staff of
the Institute of Space and Astronautical Science.



\newpage

\figcaption[figure1.ps]{An example (NGC4649)
of the single-temperature model fitting. 
Only the data for the SIS0 and GIS2 are shown. 
The solid lines show the best-fit single-temperature model. 
}

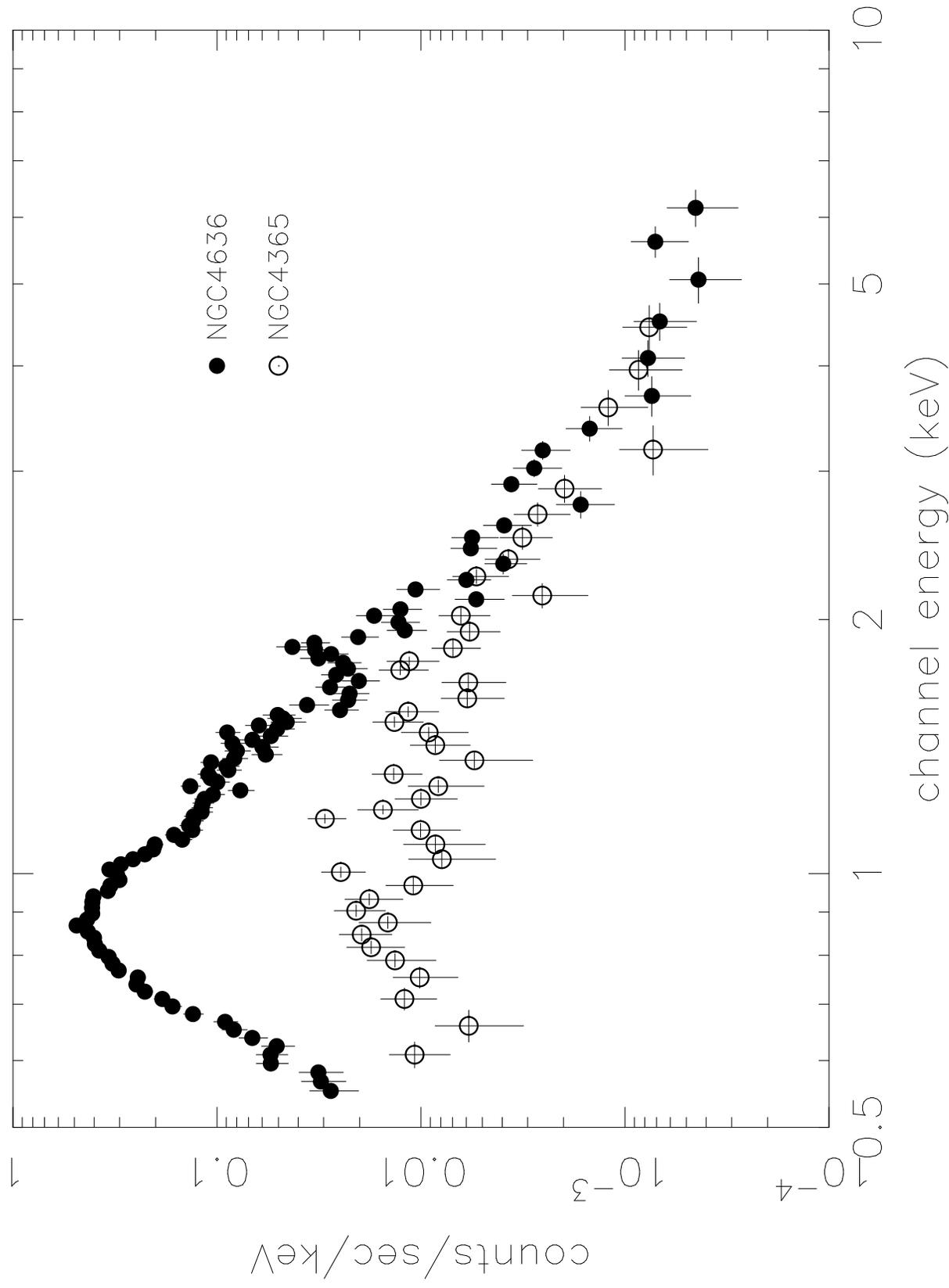
\figcaption[figure2.ps]{The SIS0 spectra of NGC4636 (filled circles) and 
NGC4365 (open circles).}

\figcaption[figure3.ps]{
The composite GIS spectra of our targets above the 4 keV band.
The solid line shows the best-fitting thermal bremsstrahlung model.
}

\figcaption[figure4.ps]{
Confidence contours at $\Delta\chi^2~=$ 2.71, 4.61 
and 9.21  for the line center energy and 
equivalent width of the Fe-K line. 
}

\figcaption[figure5a.ps]{
The relation between the best-fit parameters of the RS and 
MEKA models: (a) the column density, (b) the plasma temperature, 
(c) the metal abundance, (d)  $L_X$ of the soft component. 
In these figures, galaxies whose 
parameters were not well constrained are not shown. 
}

\figcaption[figure6.ps]{
The X-ray luminosity of the hard component vs. 
the blue band luminosity.
The downward arrow shows upper limit of NGC499.
The line shows the relation for spiral galaxies (Canizares, Fabbiano, \&
Trinchieri 1987). 
}

\figcaption[figure7.ps]{
The X-ray luminosity of the soft component vs. 
the blue band luminosity.
The line is the same as shown in Figure 6.}

\figcaption[figure8.ps]{
The volume emission measure vs temperature of the hot gas. 
Plus symbols denote clusters of galaxies. 
The line shows the best fit relation for the cluster data. 
}

\figcaption[figure9.ps]{
The velocity dispersion vs. the gas temperature. 
The solid line is given by the equation $\beta = \mu m_H \sigma^2 / kT = 1$.
The symbols are the same as in Figure 8. 
}

\figcaption[figure10.ps]{
The stellar velocity dispersion vs. $\beta_{spec}$.
}

\figcaption[figure11.ps]{
The ratio of the square root of the X-ray luminosity of 
the hot gas to the blue band luminosity vs. $k{\Delta}T$, 
where ${\Delta}T = T_1 - T_2$, 
$T_1$ is the gas temperature, and $T_2$ is the temperature predicted 
from the velocity dispersion assuming 
$\beta = \mu m_H \sigma^2 / kT_2 = 1$. 
}

\figcaption[figure12.ps]{
The metal abundance vs. the hot gas temperature.
}

\clearpage

\begin{deluxetable}{lcccccccccc} 
\tablecolumns{11}
\scriptsize
\tablewidth{0pt}
\tablecaption{Target data}
\tablehead{
\colhead{Name}	& \colhead{RA\tablenotemark{a}}	
& \colhead{DEC\tablenotemark{a}}	&\colhead{Distance{\tablenotemark{b}}}
&\colhead{$B^{T}_0$\tablenotemark{c}}	&\colhead{$M_B$\tablenotemark{d}}
&\colhead{$\log \left(\frac{\displaystyle L_B}{\displaystyle L_{\sun}}\right)$\tablenotemark{e}}	
&\colhead{Type\tablenotemark{f}}	&\colhead{$\sigma$\tablenotemark{g}}
&\colhead{$N_H$\tablenotemark{h}}	&\colhead{Group\tablenotemark{i}}
\\
&\colhead{(2000)}	&\colhead{(2000)}	&\colhead{(Mpc)}	
& \colhead{(mag)}	&\colhead{(mag)}	&\colhead{}		
&\colhead{}	&\colhead{(km/sec)}	&\colhead{($cm^{-2}$)}
}
\startdata
NGC4472(M49)	&12 29 46.5	&07 59 58	&25.8	&9.26	&-22.80	&11.28	&E2	&315	&1.6e20	&Virgo\nl
NGC4406(M86)	&12 26 11.8	&12 56 49	&25.8	&9.71	&-22.35	&11.10	&E3	&256	&2.7e20	&Virgo\nl
NGC4636		&12 42 49.8	&02 41 17	&25.8	&10.37	&-21.69	&10.84	&E0-1	&217	&1.9e20	&Virgo\nl
NGC4649(M60)	&12 43 40.3	&11 32 58	&25.8	&9.70	&-22.36	&11.11	&E2	&344	&2.4e20	&Virgo\nl
NGC499		&01 23 11.6	&33 27 36	&94.5	&12.79	&-22.09	&11.00	&S0$^-$	&235	&5.3e20	&Pisces\nl
NGC507		&01 23 40.1	&33 15 22	&94.5	&12.19	&-22.69	&11.24	&SA(r)0$^0$	&306	&5.3e20	&Pisces\nl
NGC720		&01 53 00.4	&-13 44 21	&39.6	&11.13	&-21.86	&10.91	&E5	&224	&1.4e20	&LGG38\nl
NGC4374(M84)	&12 25 03.7	&12 53 15	&25.8	&9.91	&-22.15	&11.02	&E1	&296	&2.6e20	&Virgo\nl
IC4296		&13 36 38.9	&-33 57 59	&72.8	&11.42	&-22.89	&11.32	&E	&290	&4.3e20	&HG22\nl
NGC4382(M85)	&12 25 24.7	&18 11 27	&25.8	&9.95	&-22.11	&11.01	&{\scriptsize SA(s)0$^+_{\rm pec}$}	&200	&2.7e20&Virgo\nl
IC1459		&22 57 10	&-36 27 42	&43.0	&10.83	&-22.34	&11.10	&E3-4	&316	&1.2e20	&HG15\nl
NGC4365		&12 24 27.9	&07 19 06	&25.8	&10.42	&-21.64	&10.82	&E3	&262	&1.6e20	&Virgo\nl 
\enddata
\tablenotetext{a}{2000 epoch coordinates from the Third Reference Catalogue of 
Bright Galaxies; de Vaucouleurs et al. \markcite{RC3} 1991, hereafter RC3}
\tablenotetext{b}{Distance from Donnelly, Faber, \& O'Connell 
\markcite{DFO90} 1990. 
For NGC499 and NGC507, we adopt the distance from Kim \& Fabbiano 1995}
\tablenotetext{c}{$B$ magnitude, where we list $B^T_0$ from RC3}
\tablenotetext{d}{Absolute blue magnitude calculated with the distances in 
the 4th column}
\tablenotetext{e}{Blue band luminosity in solar units, defined as 
$\log L_B = -0.4(M_B-5.41)$}
\tablenotetext{f}{Morphological type from RC3}
\tablenotetext{g}{Stellar velocity dispersion from Canizares, Fabbiano 
\& Trinchieri 1987, Tonry \& Davis \markcite{TD81} 1981, 
and McElroy \markcite{MCE95} 1995}
\tablenotetext{h}{Galactic column density from Stark et al. \markcite{SGW92} 
1992}
\tablenotetext{i}{Name of group to which the galaxy belongs. 
``Virgo'' denotes the Virgo cluster of the galaxies, 
``Pisces'' denotes the Pisces group,
``LGG'' denotes a group in Garcia \markcite{Gal93} 1993, 
and ``HG'' denotes a group in Huchra \& Geller \markcite{HG82} 1982}
\end{deluxetable}

\begin{deluxetable}{lllrlllllllll}
\scriptsize
\tablewidth{0pt}
\tablecaption{{\it ASCA} observation log}
\tablehead{
\colhead{Target}	&\colhead{Date}	&\colhead{Radius}	&
\multicolumn{2}{c}{SIS mode\tablenotemark{a}}	
&\multicolumn{4}{c}{Exposure}
&\multicolumn{4}{c}{count rate\tablenotemark{b}}\\
&\colhead{(yy/mm/dd)}&\colhead{(arcmin)}&\colhead{}&\colhead{}&
\multicolumn{4}{c}{(ksec)}&\multicolumn{4}{c}{(ksec)}\\
&\colhead{}&\colhead{}	&\colhead{}	&\colhead{}&
\colhead{SIS0}	&\colhead{SIS1}	&\colhead{GIS2}	&\colhead{GIS3}	&
\colhead{SIS0}	&\colhead{SIS1}	&\colhead{GIS2}	&\colhead{GIS3}
}
\startdata
NGC4472	&1993/07/04	&5	&B	&4	&20.1	&20.1	&21.5	&21.5	&0.39	&0.31	&0.13	&0.16\nl
NGC4406	&1993/07/03	&5	&F	&4/2	&19.7	&19.7	&20.8	&20.8	&0.34	&0.27	&0.11	&0.13\nl
NGC4636	&1993/06/22	&5	&B	&4	&38.3	&38.4	&37.6	&37.6	&0.38	&0.27	&0.090	&0.092\nl
NGC4649	&1994/01/07	&5	&F	&4/2	&38.8	&38.8	&38.8	&38.8	&0.18	&0.11	&0.070	&0.072\nl
NGC499	&1994/01/23	&1.4	&F	&2	&28.0	&28.1	&38.9	&38.8	&0.031	&0.031	&5.1e-3	&8.8e-3\nl
NGC507	&1994/01/23	&1.4	&F	&2	&28.0	&28.1	&38.9	&38.8	&0.052	&0.039	&0.020	&0.016\nl
NGC720	&1993/07/17	&3.3	&B	&4	&34.9	&35.0	&36.3	&36.3	&0.033	&0.026	&8.7e-3	&0.011\nl
NGC4374	&1993/07/04	&5	&B	&4/2	&19.3	&19.4	&20.5	&20.5	&0.048	&0.088	&0.020	&0.037\nl
IC4296	&1994/02/15	&1.8	&F	&2	&35.8	&35.8	&37.1	&37.1	&0.025	&0.018	&8.5e-3	&0.010\nl
NGC4382	&1994/05/27	&5	&F	&2/1	&33.1	&33.4	&35.3	&35.3	&0.045	&0.035	&0.011	&0.013\nl
IC1459	&1993/05/20	&3	&B	&4	&19.5	&19.5	&19.3	&19.3	&0.043	&0.032	&0.016	&0.018\nl
NGC4365	&1993/06/28	&5	&B	&4	&35.7	&35.7	&36.5	&36.5	&0.019	&0.019	&0.013	&0.015\nl 
\enddata
\tablenotetext{a}{SIS data mode. F: Faint mode, B: Bright mode, 
4: 4 CCD mode, 2: 2 CCD mode, 1: 1 CCD mode}
\tablenotetext{b}{Background subtracted value}
\end{deluxetable}

%
%

\begin{deluxetable}{lccccccccccccc}
\scriptsize
\tablewidth{0pt}
\tablecolumns{10}
\tablecaption{The Best-fit parameters and 90\% confidence errors from
the RS model fitting}
\tablehead{
\colhead{Target}  & \colhead{N$_H$}&\colhead{$kT$} & 
\colhead{Abundance\tablenotemark{a}}  	&\colhead{$VEM$\tablenotemark{b}} &
\multicolumn{2}{c}{$F_X$ in 0.5 -- 4.5 keV} &
\multicolumn{2}{c}{$L_X$ in 0.5 -- 4.5 keV} &
$\chi^2/d.o.f.$\\
&\colhead{}&&&&\colhead{soft comp.}&\colhead{hard comp.}&
\colhead{soft comp.}	&\colhead{hard comp.}\\
        & \colhead{(10$^{20}$cm$^{-2}$)}&\colhead{(keV)}	& 
\colhead{(solar)}  &\colhead{($10^{64}$cm$^{-3}$)}	&\colhead{(erg/sec)}&
\colhead{(erg/sec)}&\colhead{(erg/sec)}	&\colhead{(erg/sec)}
}
\startdata
NGC4472 & 13.4	& 0.89 	& 0.39 	&5.9	&5.5e-12	&9.0e-13	
&6.2e41		&8.4e40	&442.1/425\nl 
& (11.3--15.5) 	& (0.88--0.90) 	& (0.33--0.45)&(5.2--6.7)\nl 
NGC4406 & 10.9 	& 0.84 	& 0.31 	 	&6.3
&5.6e-12	&4.2e-13	&6.0e41		&3.9e40
&708.3/616\nl 
& (8.5--13.0) 	& (0.83--0.86) 	& (0.26--0.38)
&(5.2--7.3)\nl 
NGC4636 & 4.9 	& 0.76 	& 0.31 	 	&5.0
&5.9e-12	&3.9e-13	&5.4e41		&3.3e40
&732.9/428\nl 
& (3.6--6.3) 	& (0.75--0.77) 	& (0.28--0.35)
&(4.4--5.6)\nl 
NGC4649 & 16.1 	& 0.85 	& 0.34 	 	&2.7
&2.3e-12	&7.1e-13	&2.8e41		&6.8e40
&605.7/545\nl 
& (13.0--19.7) 	& (0.83--0.86) 	& (0.28--0.44)
&(2.2--3.3)\nl 
NGC499 & 7.8 	& 0.71 	& 0.25 		&21.3
&1.5e-12	&$<$1.2e-13	&2.0e42		&$<$1.4e41
&123.0/115\nl 
& (1.0--16.1) 	& (0.67--0.76) 	& (0.18--0.47)
&(12.8--26.6)\nl 
NGC507 & 10.7 	& 1.01 	& 0.29 	 	&33.0
&1.9e-12	&2.5e-13	&2.6e42		&3.1e41
&317.0/177\nl 
& (6.4-15.6) 	& (0.96--1.05) 	& (0.20--0.42)
&(24.5--42.6)\nl 
NGC720 & 0.0 	& 0.65 	& 0.098 	 	&1.9
&5.5e-13	&2.2e-13	&1.0e41		&4.1e40
&150.3/139\nl 
& ($<$1.3) 	& (0.61--0.68) 	& (0.069--0.14)
&(1.5--2.5)\nl 
NGC4374 & 12.1 	& 0.67 	& 0.052 	 	&3.5
&1.2e-12	&4.7e-13	&1.4e41		&4.3e40
&274.9/317\nl 
& (3.3--20.0) 	& (0.59--0.81) 	& (0.029--0.12)
&(1.7--5.5)\nl 
IC4296 & 6.8 	& 0.84 	& 1.1 		&1.1
&3.7e-13	&4.5e-13	&2.8e41		&3.1e41
&139.7/110\nl 
& ($<$27.0) 	& (0.79--0.88) 	& ($>$0.32)
&(0.18--3.7)\nl 
NGC4382 & 19.0 	& 0.28 	& 0.056 		&4.3
&3.7e-13	&5.3e-13	&7.3e40		&5.2e40
&280.4/217\nl 
& (0.43--32.5) 	& (0.22--0.33) 	& (0.029-0.21)
&(0.53--18.3)\nl 
IC1459 & 0.0 	& 0.71 	& 0.17 		&0.64
&2.2e-13	&7.2e-13	&4.8e40		&1.6e41
&84.11/109\nl 
& ($<$4.1) 	& (0.64-0.79) 	& ($>$0.065)
&(0.024--1.4)\nl 
NGC4365 & 0.0 	& 0.56 	& 0.12 		&0.11
&8.4e-14	&3.9e-13	&6.6e39		&3.1e40
&200.2/218\nl 
& ($<$32.8) 	& (0.22-0.72) 	& ($>$0.016)
&(0.0022--8.7)\nl
\enddata
\tablenotetext{a}{We assumed the solar ratio of number density of iron to 
that of hydrogen is $4.68 {\times} 10^{-5}$ (Anders and Grevesse 1989)}
\tablenotetext{b}{Volume Emission Measure of the soft component}
\end{deluxetable}

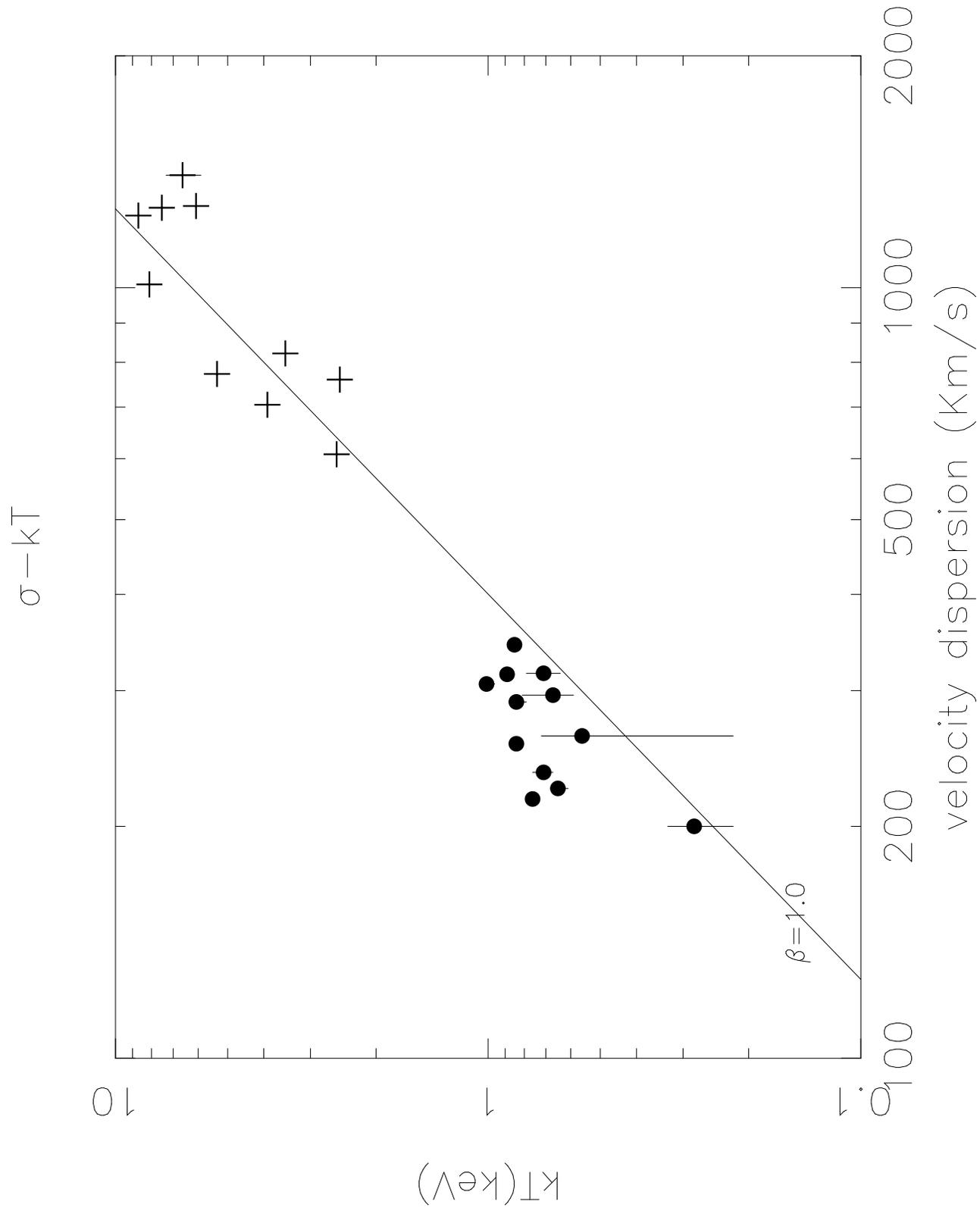
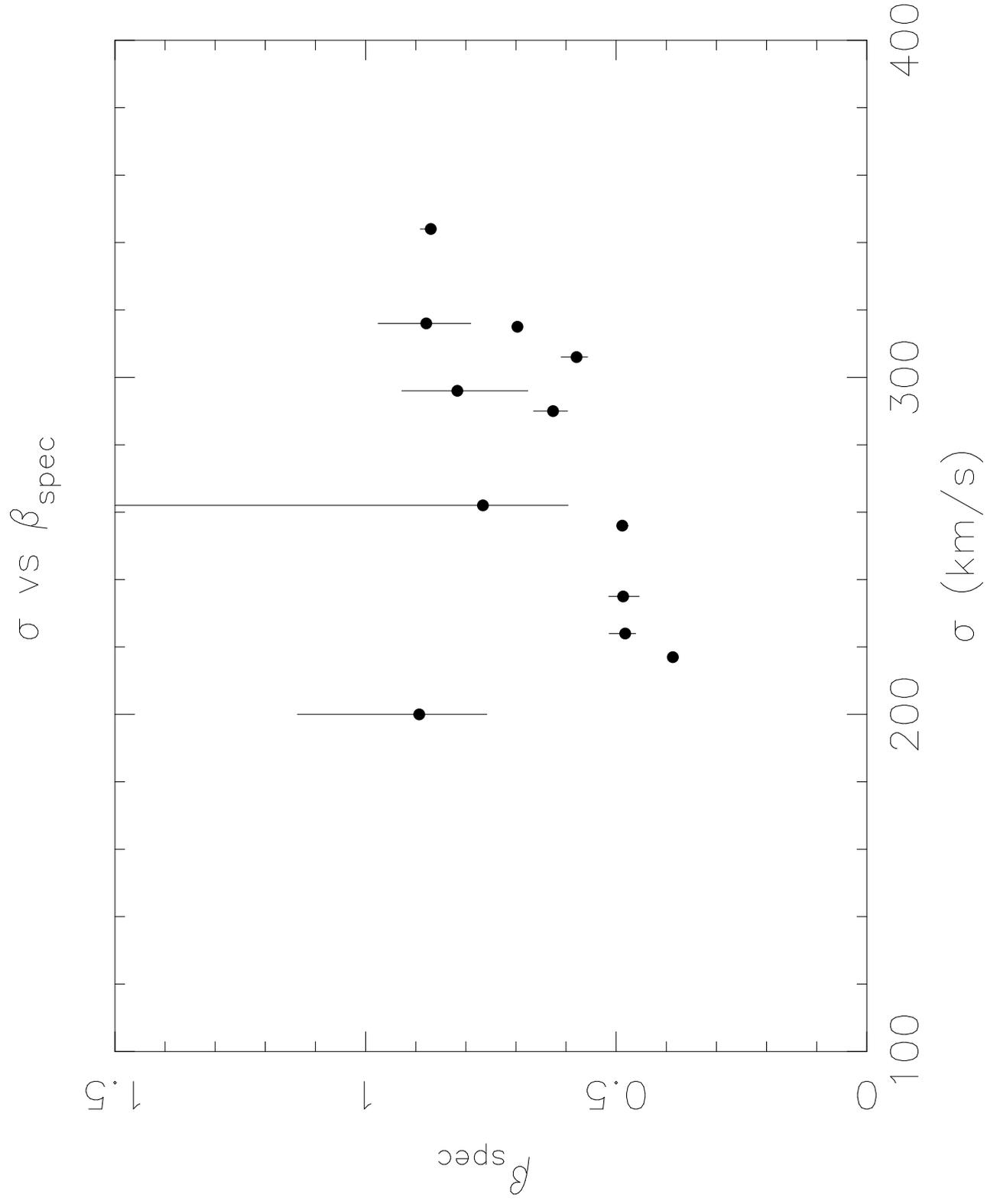
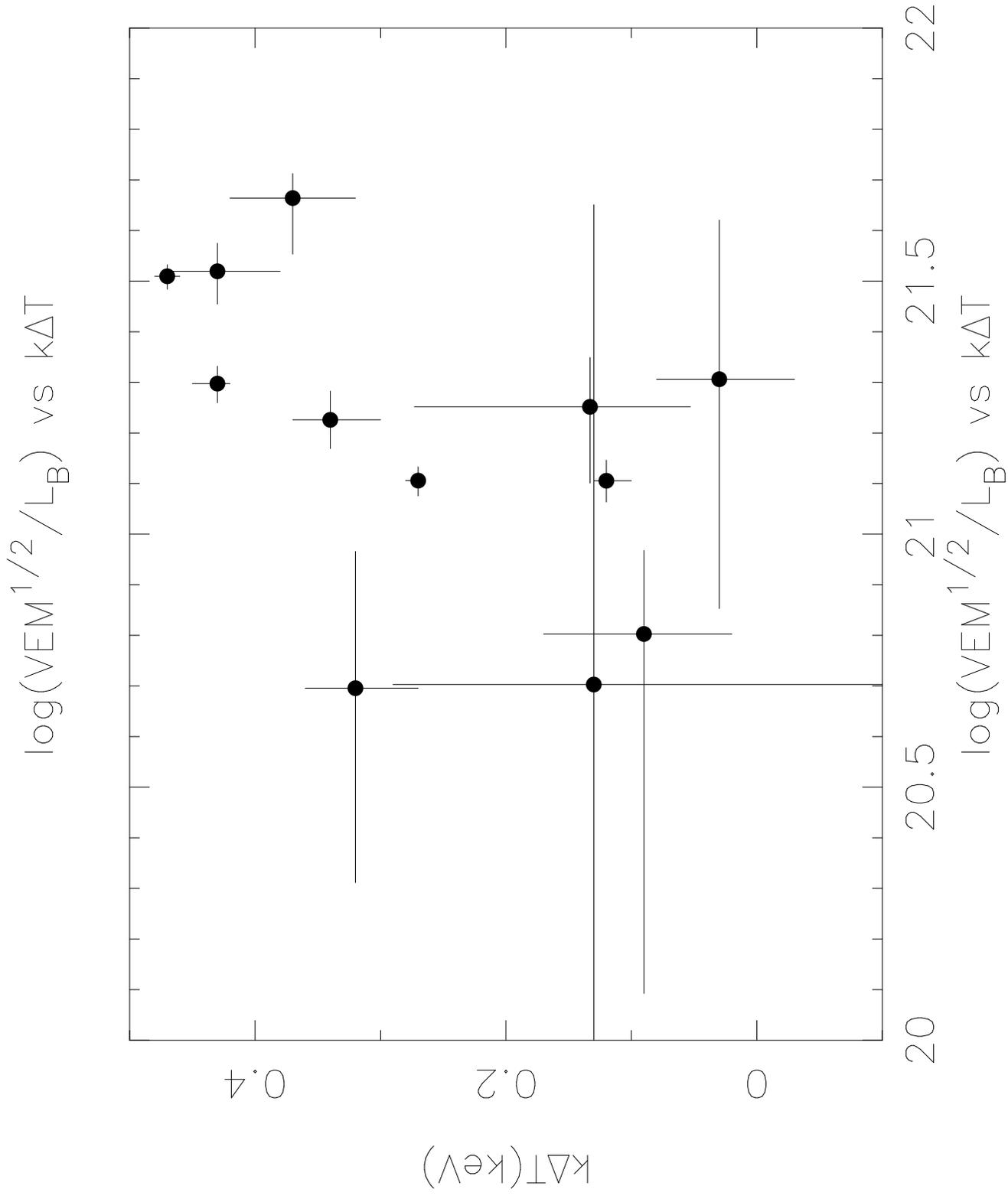
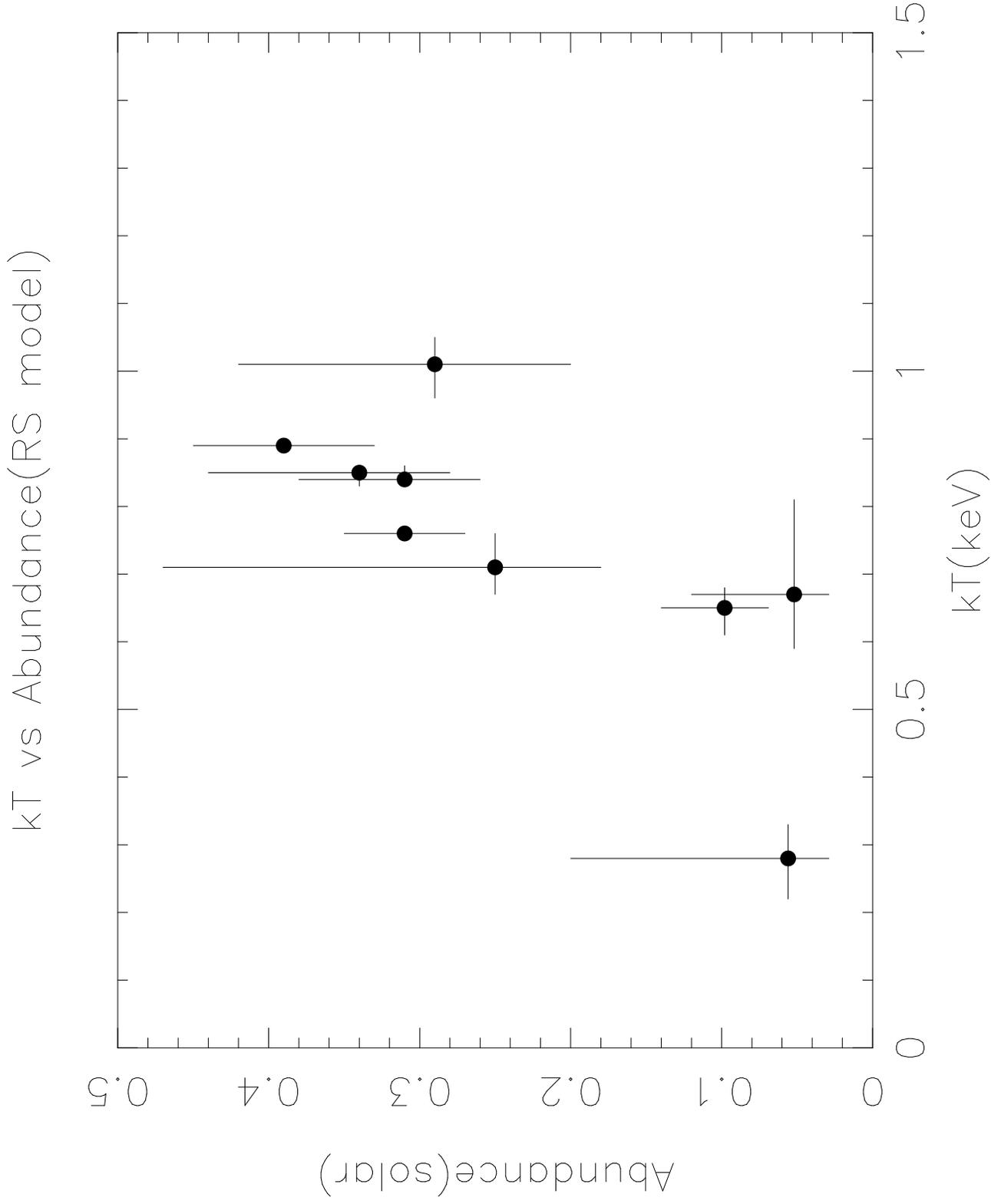
\end{document}